# COMMUNICATION

# Multi-objective Bayesian Optimization with Human-in-the-Loop for Flexible Neuromorphic Electronics Fabrication

Benius Dunn,[a] Javier Meza-Arroyo,[a] Armi Tiihonen,[b] Mark Lee [c] and Julia W. P. Hsu *[ac]

Neuromorphic computing hardware enables edge computing and can be implemented in flexible electronics for novel applications. Metal oxide materials are promising candidates for fabricating flexible neuromorphic electronics, but suffer from processing constraints due to the incompatibilities between oxides and polymer substrates. In this work, we use photonic curing to fabricate flexible metal-insulator-metal capacitors with solution-processible aluminum oxide dielectric tailored for neuromorphic applications. Because photonic curing outcomes depend on many input parameters, identifying an optimal processing condition through a traditional grid-search approach is unfeasible. Here, we apply multi-objective Bayesian optimization (MOBO) to determine photonic curing conditions that optimize the trade-off between desired electrical properties of large capacitance-frequency dispersion and low leakage current. Furthermore, we develop a human-in-the-loop (HITL) framework for incorporating failed experiments into the MOBO machine learning workflow, demonstrating that this framework accelerates optimization by reducing the number of experimental rounds required. Once optimization is concluded, we analyze different Pareto-optimal conditions to tune the dielectric's properties and provide insight into the importance of different inputs through Shapley Additive exPlanations analysis. The demonstrated framework of combining MOBO with HITL feedback can be adapted to a wide range of multi-objective experimental problems that have interconnected inputs and high experimental failure rates to generate usable results for machine learning models.

**New Concepts**

This work introduces a framework for integrating experimental failures into machine-learning-driven multi-objective optimization of physical processing problems. Previous approaches in this area often use binary classification models to identify experimental regions of high feasibility, a method that fails to capture distinct failure modes or distinguish between experimental outcomes that are near or far from the feasible region. By incorporating quick, subjective input from an experienced researcher into the machine learning process, the number of experiments required to achieve convergence is significantly reduced, thereby saving time and resources in the laboratory. This framework is demonstrated through the fabrication of dielectrics for flexible neuromorphic electronics using photonic curing, a multidimensional problem with many interconnected input parameters. Human observations can be quickly conducted and provide information on different failure modes and the severity of failure, which would be difficult to automate. By directing the optimizer to avoid unpromising regions, a significant improvement in device yield is achieved compared to traditional multi-objective optimization. This framework is applicable to any experimental workflow where failed or borderline outcomes provide valuable information, often encountered in materials science and device fabrication where results often span a continuous spectrum rather than a simple pass-or-fail outcome.

## Introduction

With the large volume of data being processed in today's technological landscape, bottlenecks in computational speed and power consumption in traditional computer architectures, i.e., von Neumann architecture, are of increasing concern. Neuromorphic computing seeks to overcome these obstacles by mimicking the highly efficient biological processes of the human brain, enabling in-memory computing and in-sensor processing at the edge.[1,2] Further, neuromorphic hardware can be expanded to the realm of flexible electronics for applications such as wearable biomedical devices and conformal sensors.[3,4]

[a.] Department of Materials Science and Engineering, The University of Texas at Dallas, Richardson, Texas 75080, USA
[b.] Department of Applied Physics, Aalto University, Espoo, Finland
[c.] Department of Physics, The University of Texas at Dallas, Richardson, Texas 75080, USA
† Electronic supplementary information (ESI) available. See DOI: 10.1039/x0xx00000x

For neuromorphic devices to be used in temporal signal processing, they must exhibit short-term synaptic plasticity (STP) behavior, with a decreasing current response over time or 'fading memory' behavior. The STP characteristics have been widely seen in metal oxide memristors[2–5] and thin-film transistors[6–8]. Devices exhibiting STP often show hysteretic behavior, attributed to different time scales associated with defects at the semiconductor/dielectric interface or with mobile species in the dielectric, most often in the form of oxygen vacancies.[5–7] Migration of mobile oxygen vacancies under electrical bias leads to a time-dependent buildup of a positive charge gradient within the dielectric, followed by a time-dependent relaxation as mobile vacancies diffuse back to their equilibrium distribution upon removing the electrical bias, resulting in hysteresis in neuromorphic transistor devices.[7] Literature has also shown that the hysteretic transistor behavior necessary for STP is often accompanied by frequency dependence of capacitance, i.e., capacitance-frequency (C-f) dispersion, in metal-insulator-metal (MIM) structures with the dielectric.[8–10]

Metal oxides have shown promise for building flexible neuromorphic devices. However, there are challenges associated with using solution-processed metal oxides on flexible polymer substrates. These include high processing temperatures and a mismatch between the coefficients of thermal expansion of metal oxides and substrates, leading to processing restrictions to prevent mechanical failures. One approach to alleviate these problems is to convert the metal oxide films using photonic curing, which uses high-intensity broadband (200 – 1500 nm) light from a xenon flash lamp to anneal a film in a millisecond timescale (< 20 ms) while leaving underlying substrates close to ambient temperature.[11,12] Several interconnected input parameters control the photonic curing process, and optimizing them requires navigating a complex multi-dimensional input space, which is impractical with traditional experimental grid-search methods.

Bayesian optimization (BO) is a popular machine learning technique for optimizing black-box functions and has been used in materials science problems.[13–18] By mapping input-output relationships via a surrogate model and sequential active learning that balances exploiting known relationships with exploring the areas of uncertainty, BO is an ideal method for optimizing processing problems with multiple input variables. While the original intent of BO is to find the global optimum of a single objective function over a pre-defined input space, this method can also be extended to optimize multiple objective functions simultaneously, known as multi-objective Bayesian optimization (MOBO).[19–21] In most multi-objective problems, different objectives compete, so the global optimal values of each cannot be achieved simultaneously. The general goal of MOBO is thus to find a Pareto frontier of optimal conditions, representing the set of conditions where one objective cannot be improved without degrading another.[22]

Based on the literature of hysteretic transistors,[8–10] we use large C-f dispersion, the ratio of low to high-frequency capacitance, in a MIM capacitor as a metric for a desirable electrical property for the metal oxide dielectric. Additionally, minimal leakage current in a dielectric is necessary for practical transistor function, thus giving us a two-objective optimization problem. These two objectives are inherently competing, as minimizing leakage current requires a denser, more stoichiometric dielectric with fewer structural defects, whereas large C-f dispersion relies on defect states. This trade-off highlights the need for a multi-objective optimization approach. In this work, we apply MOBO to convert sol-gel $Al_2O_3$ dielectric using photonic curing for MIM structures on flexible polyethylene terephthalate (PET) substrates.

Unlike synthetic data, real experiments can produce results that fail, which cannot be used to train a machine learning surrogate model. To remedy this issue, we implement a human-in-the-loop (HITL) to incorporate unsuccessful photonic curing results into the MOBO workflow. Observations by human scientists provide fast and valuable information that is difficult for a machine to replicate, which would require defining success versus failure and adding another layer of characterization. Previous work has demonstrated the merits and methodology of incorporating subjective human inputs into BO.[23,24] After incorporating HITL, we successfully inform the machine learning model about the input space with a higher probability of producing working devices while continuing to optimize electrical characteristics. Through simultaneously mitigating leakage current and maximizing C-f dispersion, we obtain a set of Pareto-optimal conditions that enable us to tune MIM characteristics for neuromorphic computing applications. We also apply SHapley Additive exPlanations (SHAP) analysis to understand the contributions of individual input parameters on each objective, enabling better insight into the physical processes affecting photonic curing. This optimization approach, which incorporates HITL knowledge, is applicable to a broad range of complex experimental and processing problems that are hindered by failed results unusable for training machine learning surrogate models, or that can benefit from incorporating domain expertise from humans.

## Experimental Methods

### $Al_2O_3$ Sol-gel Film Preparation

The precursor solution for $Al_2O_3$ films is prepared by dissolving 0.4 M $Al(NO_3)_3 \cdot 9H_2O$ in 2-methoxyethanol (2-MOE) and stirring with the vial uncapped at 80°C for 18 hours. Evaporated solvent is replaced with fresh 2-MOE, and the solution is filtered through a 0.22 μm PTFE filter immediately before use. 100-μm-thick heat-stabilized PET substrates (KODAK 6RF1-502) are cleaned in an ultrasonic bath with isopropyl alcohol for 10 minutes, then rinsed with deionized (DI) water and dried with a $N_2$ stream. 100-nm-thick Al bottom gate contacts are deposited by thermal evaporation through a shadow mask before dielectric deposition. These substrates are cut into 25 mm square pieces and treated with UV-ozone for 20 minutes to clean off contaminants and improve solution wetting. The aluminum nitrate precursor solution is spin-coated at 3000 rpm for 30 seconds and then dried for 2 minutes at 90°C on a hot plate.

## Photonic Curing

Photonic curing is performed with a PulseForge Invent[25] equipped with three 950 V lamp drivers, a 1.5 kW power supply, and a 20 mm diameter × 150 mm length xenon flash lamp. The photonic curing system includes five control parameters: (1) radiant energy, also known as pulse fluence, is the amount of energy per pulse delivered to the sample by the flash lamp per unit area, in units of $J/cm^2$; (2) pulse count, the number of flash lamp pulses delivered to the sample; (3) pulse length, the time over which the flash lamp is on, in units of milliseconds; (4) number of micropulses, which divides a single pulse into sub-pulses; and (5) duty cycle, the percentage of the entire pulse length for which the flash lamp is on. The aluminum nitrate precursor films are placed on the sample stage at 6 mm below the lamp. A 0.005 in. Invar shadow mask is placed on top of the sample to define the light exposure areas, and both are held in place by steel slats with magnets. After photonic curing, the patterns are developed by submerging the sample in a weak acid solution consisting of a 15:5:1 volume ratio of methanol:DI water:glacial acetic acid for 5 seconds to remove the unexposed parts, followed by rinsing with DI water and drying with a $N_2$ stream.

## MIM Fabrication and Characterization

After photonic curing of the dielectric, Al metal top contacts are deposited by the same procedure as bottom gate contacts. MIM capacitors with areas defined by the top contact range from $4 \times 10^{-4}$ $cm^2$ to $64 \times 10^{-4}$ $cm^2$ in size. C-f measurements are performed using an Agilent 4284A Precision LCR Meter over a range of $10^2$ to $10^6$ Hz with 0 V DC bias and 50 mV AC amplitude. Leakage current-voltage (I-V) measurements are performed using a Keithley 4200A Semiconductor Characterization System by DC voltage sweep from -5 V to 5 V and back to -5 V.

## Machine Learning Methods

### Multi-objective Bayesian Optimization

To achieve desirable neuromorphic electronics, the objectives of MOBO are to maximize the C-f dispersion (representing the dielectric property that promotes the STP behavior) and minimize leakage current (representing the gate dielectric requirement for a functioning transistor) of the MIM devices. The electric characteristics depend on five photonic curing parameters (defined in 'Photonic Curing'): radiant energy, number of pulses, pulse length, number of micropulses, and duty cycle. The search space encompasses over 4 million distinct combinations of parameter values for the photonic curing process (Table S1, ESI†), rendering it infeasible to explore manually through laboratory experiments. This motivates the use of MOBO, with these five parameters serving as input variables.

An initial set of 30 photonic curing conditions (Table S2, ESI†) is generated using the pseudorandom sampling method of Latin hypercube sampling (LHS) as the first step of the MOBO workflow (Fig. 1(a)). This initial sampling across the entire input space (distribution of LHS conditions is detailed in Fig. S1, ESI†) is used to convert $Al_2O_3$ dielectric films for MIM capacitors (Fig. 1(b)). Full MIM devices are then characterized by C-f and I-V measurements (Fig. 1(c)). The electrical results are used to train two separate Gaussian process regression (GPR) models as surrogates for the C-f dispersion and leakage current (Fig. 1(d)). To perform MOBO and fit GPR models, a quantitative numeric metric needs to be defined for both of these target properties. First, we define a proxy variable for C-f dispersion using the ratio of the areal capacitance value at $10^2$ Hz to that at $10^6$ Hz, $C_{100Hz}/C_{1MHz}$, of MIM capacitors. This should be maximized. Second, we define a proxy variable for gate leakage current, $I_{leakage}$, as the mean areal leakage current at ±5 V bias. $I_{leakage}$ varies over several orders of magnitude; thus, we convert it to logarithmic values to ensure better performance with the GPR model. We use the absolute value, $|\log(I_{leakage})|$, to transform the optimization into a maximization problem for both target variables.

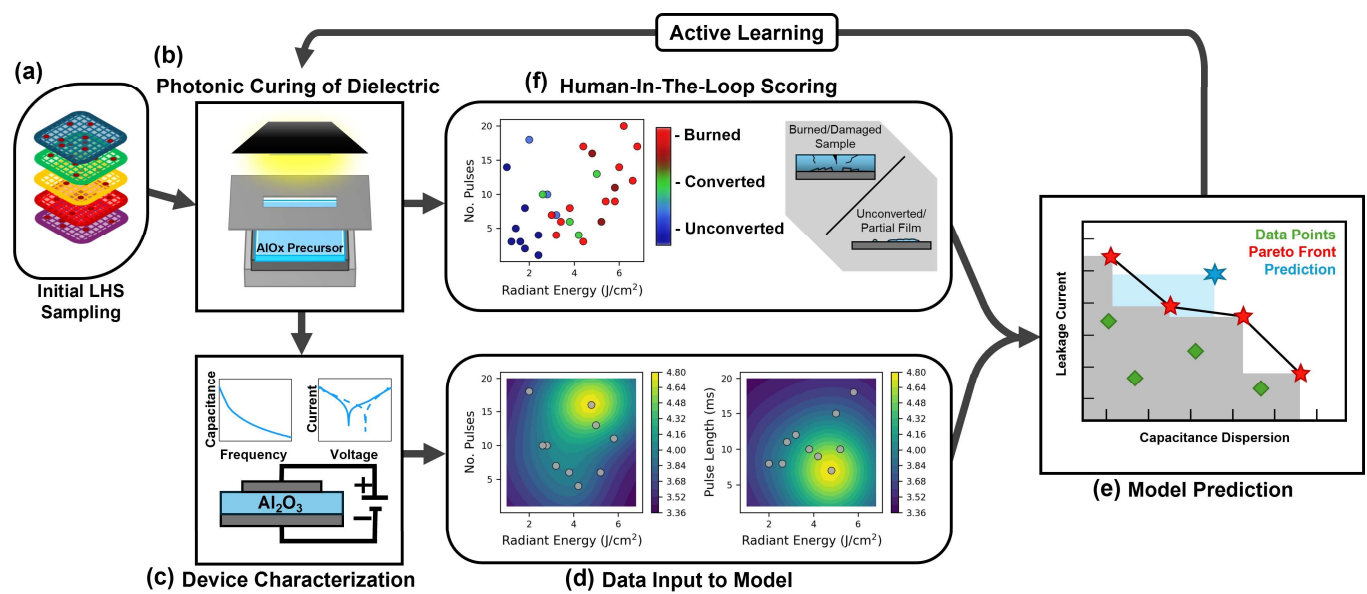

**Fig 1.** Workflow for multi-objective optimization of photonic curing conditions with human-in-the-loop incorporation. (a) Initial LHS sampling for photonic curing conditions, (b) dielectric fabrication with photonic curing, (c) characterization of MIM devices for C-f dispersion and leakage current, (d) training GPR models on outputs of electrical measurements, (e) construction of Pareto frontier and prediction of next batch inputs for photonic curing, and (f) human-in-the-loop scoring of film conversion. This process is reiterated, starting back at (b) with all data collected from previous rounds.

After building the initial surrogate models (Fig. S2, ESI†), an acquisition function is used in the subsequent rounds to generate the next set of photonic curing conditions to test, as depicted in Fig. 1(e). These conditions are used to fabricate new devices in each subsequent round, the electrical results of which are fed into the surrogate models. This active learning process is iterated, with a new Pareto frontier of the model and measured data being generated each round. Optimization is deemed completed when all new electrical results fall within ±1 standard deviation of the model-predicted values. Rounds 1 through 5 of MOBO batch picking were conducted using the parallel expected hypervolume improvement (qEHVI) acquisition function.

In order to obtain a more diverse array of model predictions along the Pareto frontier, later rounds implement a new MOBO batch picking strategy we call Pareto-UCB. We adapt and modify the method from Lukovic et al.[26] by first evaluating the acquisition function for each GPR model, then calculating the Pareto frontier of these acquisition values. We then select the next batch of conditions to test by choosing points along this newly calculated Pareto frontier of acquisition function values.

While Lukovic et al. use the mean posterior function of the GPR as the acquisition function, $acq_{mean} = \mu$, we employ upper confidence bounds (UCB) as our acquisition function, $acq_{UCB} = \mu + \beta^{1/2}\sigma$, to increase exploration of unknown areas of the parameter space. $\mu$ is the GPR mean posterior model values, $\sigma$ is the GPR posterior uncertainty, and $\beta$ is the hyperparameter that controls exploration vs. exploitation. We choose a $\beta$ value of 2 in this work to encourage exploration of the parameter space. We calculate the Pareto frontier of non-dominated UCB values and select the next batch of conditions to test by greedily maximizing the hypervolume enclosed by this frontier.

The GPR models, optimization with qEHVI acquisition function, and hypervolume calculation for Pareto-UCB batch picking are implemented in Python using the BoTorch package.[27] GPR models use the ARD Matern 5/2 kernel, with kernel length scales fitted independently for each input variable in each model, hence allowing each input variable to impact each objective ($C_{100Hz}/C_{1MHz}$, $|\log(I_{leakage})|$) to a different extent.

**Incorporating Human Input**

A challenge revealed in this study is process failures, which are often encountered in real experiments. Not all photonic curing conditions yield a converted film that can produce measurable devices; some are partially converted with non-uniform coverage, while others result in burned films or damaged substrates (Fig. S3, ESI†). Consequently, such failed devices generate no $C_{100Hz}/C_{1MHz}$ or $|\log(I_{leakage})|$ data. When only the results from functional devices are fed into the surrogate regression models for choosing the next test conditions, the BO algorithm is prone to continuously suggest processing conditions in regions that produce failed experiments, which it sees as unexplored areas. In cases where certain regions of the domain clearly outperform others, one possible solution is to shrink the domain space to focus on promising regions.[28] However, in our experiment, this is infeasible due to the highly interconnected nature of the photonic curing input parameters. A set constant value of one input parameter can produce films ranging from unconverted to burned by altering the other four input parameters. While it has been shown that constraining input space as a function of multiple input parameters is possible,[29] doing so requires knowledge of which input combinations will produce failed outcomes. As such, we need a way to inform the model which areas of the input space can successfully produce a film without prior knowledge, based on human-observed conversion outcome (Fig. 1(f)).

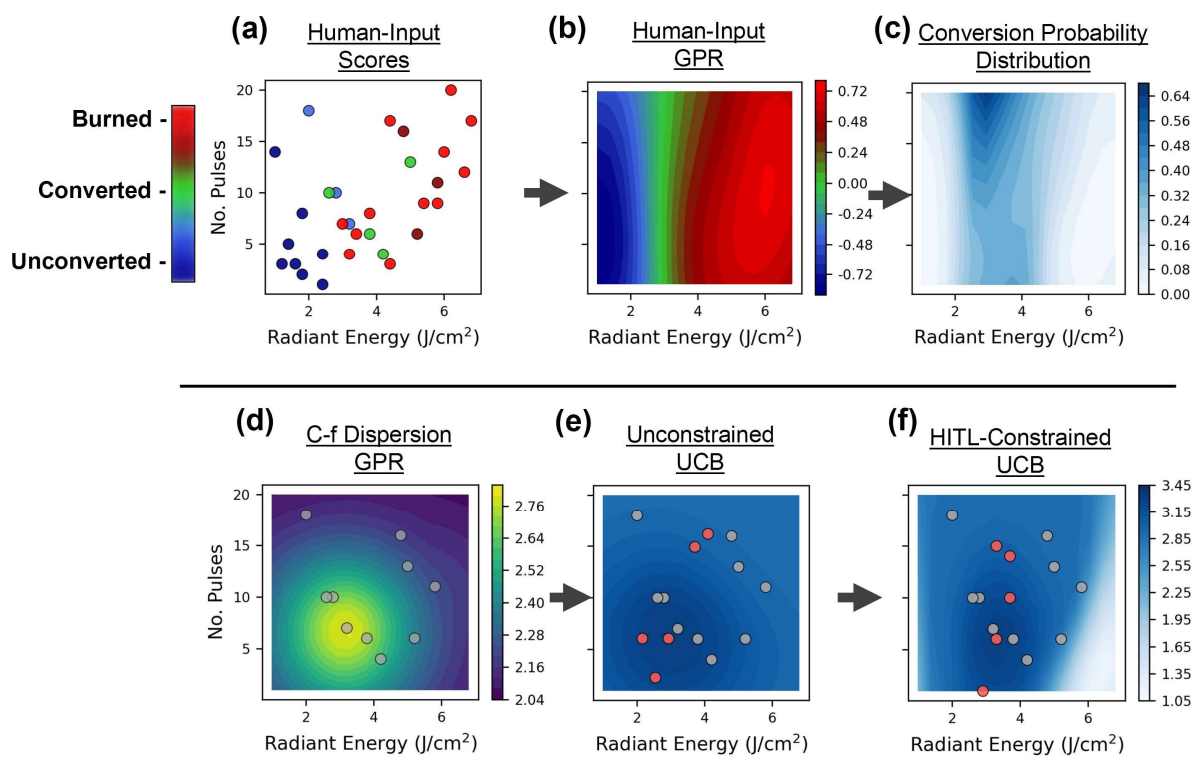

**Fig. 2.** Framework for HITL incorporated MOBO for the LHS dataset, depicted for one pair of photonic curing input parameters, radiant energy vs. number of pulses. (a) All 30 initial LHS conditions are color-coded by human observations of the films after photonic curing. (b) Mean posterior value ($\mu_{conv}$) of the GPR model trained on the human-observation scores shown in (a). (c) Probability Distribution, $P_{constraint}$, with 1 representing a high probability of successfully producing a functional device and 0 a low probability, calculated from HITL GPR using a Gaussian transformation. (d) GPR model trained on measured $C_{100Hz}/C_{1MHz}$ of successfully converted MIM devices. (e) Upper confidence bounds acquisition function of the $C_{100Hz}/C_{1MHz}$ model, unconstrained by $P_{constraint}$. (f) Upper confidence bounds acquisition function of the $C_{100Hz}/C_{1MHz}$ model after constraining with $P_{constraint}$ shown in (c). Gray dots represent previously measured points that the model is trained on, while red dots represent the conditions selected by the model for testing in the next round of MOBO.

One approach is to constrain the acquisition function for next-batch BO picks using a binary classification model that predicts whether a given input condition will produce a successful (model value of 1) or failed (model value of 0) outcome, which can be based on either human observation or automated characterization.[30–32] However, this approach is limited in that it does not give a physical model of different modes of failure, i.e., burning versus under-conversion in our experiment. Additionally, the methodology of binary classification cannot incorporate knowledge about conditions that give outcomes closer to success, such as partially burned and partially under-converted conditions. Following this concept, in order to constrain our MOBO in a way that accurately captures the different failure mechanisms and distinguishes between clear failures and partial failures close to success, we assign a numerical score to each photonic curing conversion outcome based on human observation depicted in Fig. 2(a). We assign unconverted = -1.0, partially converted = -0.5, converted = 0, partially burned = +0.5, and burned = +1.0. These values are then used to train a separate GPR model to represent film conversion outcome across the input space (Fig. 2(b)). This film conversion GPR is then used to calculate a probability distribution of yielding a functional film for a given photonic curing condition through a nonlinear transformation, similar to the methodology from Tiihonen et al.[23] Specifically, we use a Gaussian function,

$$P_{constraint} = e^{-\frac{1}{2}\left(\frac{\mu_{conv}}{\tau}\right)^2}$$

where $P_{constraint}$ (Fig. 2(c)) is the probability distribution of successful film conversion, $\mu_{conv}$ is the value of the film conversion GPR posterior mean model, and $\tau$ acts effectively as a hyperparameter for weighing the relative importance of HITL observations versus MIM electrical characterizations for the next batch picks. Because we want to increase device yield by strongly avoiding regions that do not produce usable films, we chose a small $\tau$ value of 0.2, ensuring that fully unsuccessful conditions ($\mu_{conv} = \pm 1$) result in $P_{constraint} \approx 0$ and will not be repeated in future batches.

The UCB acquisition function of each objective is then weighted by the probability of successful film conversion, $acq_{constrained} = acq_{UCB} \cdot P_{constraint}$, prior to each round of MOBO batch picking to favor the search space that is more likely to produce measurable devices and avoid the search space likely to produce process failures. Fig. 2(d) shows this process for a single input pair, radiant energy vs. number of pulses, and a single objective, $C_{100Hz}/C_{1MHz}$, using the GPR model of C-f dispersion. When the UCB acquisition function, Fig. 2(e), is constrained by $P_{constraint}$, the new HITL-constrained acquisition function, Fig. 2(f), is suppressed in areas that are less likely to produce a functional film (high radiant energy in this example), and the next batch of conditions generated by the model is shifted accordingly. Fig. 2 depicts these for the single pair of photonic curing inputs. Representations for all pairs of photonic curing inputs are included in Fig. S4 (ESI†).

## Results and discussion

Of the 30 initial LHS photonic curing conditions, only 10 yield functional MIM devices. Their $C_{100Hz}/C_{1MHz}$ and $|\log(I_{leakage})|$ values are used to build the first surrogate model. For the next two rounds of active learning, 15 photonic curing conditions are generated with MOBO and tested. But the failure rate of these selections is very high, with only 5 conditions producing measurable devices. Thus, we implement the HITL methodology discussed in Section 3.2 in subsequent rounds. Five additional active learning rounds are conducted. Electrical data from samples made in the fifth round fall within ± 1 standard deviation of the model-predicted Pareto frontier. We therefore conclude that the MOBO with HITL framework has successfully identified the Pareto frontier to within experimental uncertainty.

Fig. 3(a) shows the Pareto frontier of the final GPR models of $C_{100Hz}/C_{1MHz}$ and $|\log(I_{leakage})|$ (blue open squares), as well as the data points for all photonic curing conditions that lead to functional devices (green diamonds) and the devices fabricated using Pareto optimal conditions (red diamonds with error bars showing standard deviation). Experimental values and error are determined from measuring 5 different MIM devices on a given sample. Based on prior experience, $|\log(I_{leakage})| \geq 4$ is needed to obtain functional thin-film transistors on top of a gate dielectric. Fig. 3(b) shows a subset of the Pareto frontier in this leakage current range of the photonically cured $Al_2O_3$ dielectric (standard deviation of the GPR model represented by the shaded region).

Also shown in Fig. 3(b) is a section of the model Pareto frontier generated from the initial LHS conditions (black open squares), demonstrating that MOBO simultaneously improves both objectives, with the most significant improvement seen in $|\log(I_{leakage})|$. The evolution of the dominated hypervolume of the Pareto optimal conditions as function of active learning is shown in Fig. S5 (ESI†). The accuracy of the final GPR models for both objectives is discussed in the ESI† section 'Dominated

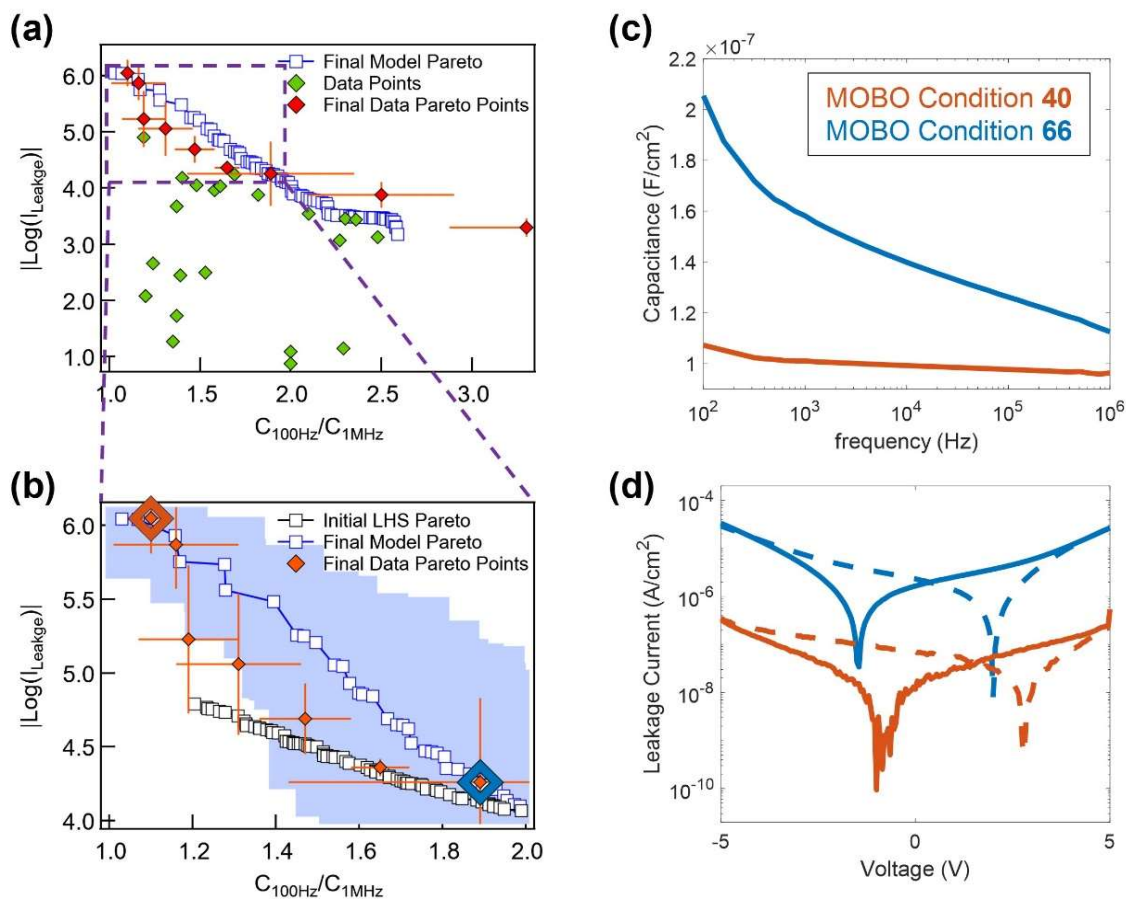

**Fig. 3.** (a) All experimental data and the final GPR modeled Pareto front plotted for $|\log(I_{leakage})|$ vs C-f dispersion ($C_{100Hz}/C_{1MHz}$). (b) Expanded view of region feasible for thin film transistor fabrication, comparing modeled Pareto frontier from initial LHS (black open squares) vs final (blue open squares) data. The blue shaded region represents the uncertainty of the modeled Pareto front, defined by ± one standard deviation of each point on the Pareto front. The red lines are standard deviations of experimental results from 5 MIM devices. In the final data Pareto front, two contrasting MIM behaviors, MOBO #40 (orange) and #66 (blue), are highlighted. (c) C-f and (d) $I_{leakage}$ results for MIM devices fabricated with MOBO #40 (orange) and #66 (blue). Solid lines in (d) represent forward I-V sweeps (-5V to +5V) and dashed lines represent reverse sweeps (+5V to -5V).

hypervolume evolution & final GPR models' accuracy' and Fig. S6.

For Pareto optimal conditions with $|\log(I_{leakage})| \geq 4$, there are several experimental photonic curing conditions with distinctly different dielectric characteristics. The two extreme electrical characteristics of MIMs made by Pareto-optimal processing conditions are MOBO conditions #40 and #66, which are highlighted in Fig. 3(c) and 3(d). Table 1 shows the photonic curing input parameters for these two conditions along with their respective objective values. Condition #40 shows the lowest leakage current (highest $|\log(I_{leakage})|$) and smallest $C_{100Hz}/C_{1MHz}$ value, while condition #66 shows the highest leakage current (lowest $|\log(I_{leakage})|$) and largest $C_{100Hz}/C_{1MHz}$ value. More in-depth analysis on the physical and chemical properties of samples made with these two conditions is presented in the ESI† section 'Physical and chemical property comparison of dielectrics from MOBO conditions #40 and #66' and Fig. S7 and S8.

To demonstrate the effectiveness of our HITL methodology in improving device yield and thereby accelerating MOBO, we revisit our initial 30 LHS conditions, of which only 10 led to measurable devices, as depicted in Fig. 4(a). The initial two rounds, based solely on device results, suggest 15 conditions, of which only five successfully produce functional devices (Fig. 4(b)). To test HITL methodology, we restart MOBO using the 30 LHS results, now incorporating the unsuccessful conditions via the HITL method for two rounds of optimization, to make a direct comparison with results based solely on electrical characterization. For the two rounds of HITL + MOBO, the device yield is 9 out of 10 conditions, which is a marked

**Table 1** Photonic curing input parameters and device output results for MOBO conditions #40 and #66

| | Input Values | | | | | Objective Values | |
|---|---|---|---|---|---|---|---|
| Condition # | Radiant Energy (J/cm$^2$) | Pulse Count | Pulse Length (ms) | Micropulse Count | Duty Cycle (%) | $C_{100Hz}/C_{1MHz}$ | $|\log(I_{leakage})|$ |
| 40 | 4.5 | 16 | 7 | 23 | 70 | 1.10 ± 0.04 | 6.05 ± 0.24 |
| 66 | 4.8 | 15 | 16 | 20 | 70 | 1.89 ± 0.46 | 4.26 ± 0.57 |

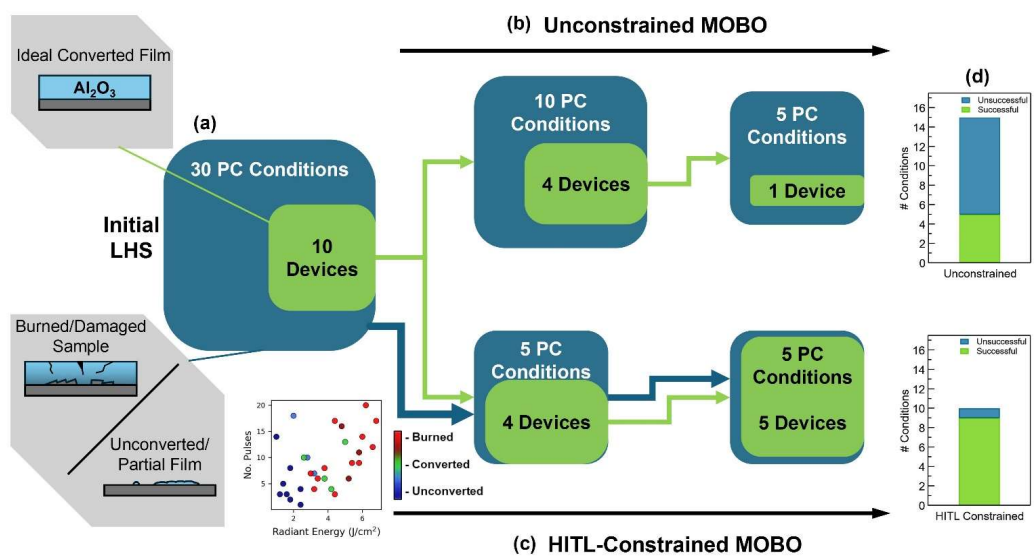

**Fig. 4.** (a) Device yield outcomes for initial LHS. Two rounds of MOBO using (b) unconstrained acquisition function without HITL versus (c) HITL-constrained acquisition function described in Figure 2. (d) Comparison of outcomes for both methodologies. Green represents successfully converted PC conditions that produce functional MIM devices. Blue represents unsuccessful PC conditions that cannot be used for device fabrication. (PC: Photonic Curing)

improvement (Fig. 4(c)). This demonstrates that HITL incorporation can substantially increase the number of data points successfully generated from each round, even from the initial sampling stage. The final results of this comparison are depicted by the yield shown in Fig. 4(d). Thus, incorporating HITL will lead to faster convergence of MOBO, requiring fewer rounds of experiments and saving time and experimental resources. Summary of the methodology used in each round of MOBO is presented in Table S4 (ESI†), while photonic curing input conditions and the processing outcomes with and without HITL are compared in Table S5 (ESI†).

To better understand the relationship between photonic curing variables and dielectric behavior, SHAP analysis is conducted on the GPR models for $C_{100Hz}/C_{1MHz}$ and $|\log(I_{leakage})|$. SHAP is a unified approach based on game theory to interpret machine learning model outcomes.[33] SHAP values give a comparative measure of the magnitude of change each input variable has on the model outcome, as well as whether it has a positive or negative effect on the outcome. For example, if an input has high SHAP values for high input values and low SHAP values for low input values, that means that the model output has a strong positive correlation with that particular input. SHAP value trends in Fig. 5 illustrate the competing nature of $C_{100Hz}/C_{1MHz}$ and $|\log(I_{leakage})|$ objectives. Fig. 5(a) shows that modeled $C_{100Hz}/C_{1MHz}$ increases with lower radiant energy, lower pulse count, and longer pulse length. Conversely, Fig. 5(b) shows that modeled $|\log(I_{leakage})|$ increases with higher radiant energy, higher pulse count, and shorter pulse length. Both objectives exhibit a small but positive relation with micropulse count. While $|\log(I_{leakage})|$ has a positive relation with duty cycle, $C_{100Hz}/C_{1MHz}$ is insensitive to duty cycle. SHAP value dependence on individual photonic curing parameters is further detailed in ESI† section 'SHAP Analysis Scatter Plots for Individual Input Parameters' and Fig. S9 and S10. The results of SHAP analysis on the GPR models reveal how different photonic curing parameters can have conflicting effects on the two objectives, as well as the magnitude of their respective effects. The photonic curing parameters and MIM characteristics of our two extreme MOBO conditions, #40 and #66, further confirm the SHAP analysis. The most significant difference in the input parameters for these two conditions (Table 1) is pulse length, with 7 ms for #40, which has a low leakage and small C-f dispersion, and 16 ms for #66, which has a high leakage and large C-f dispersion.

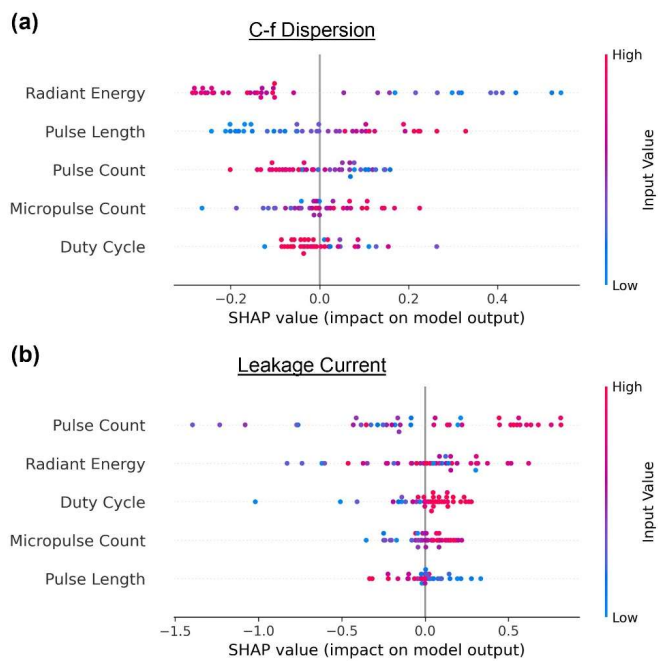

**Fig. 5.** SHAP analysis of GPR models for (a) C-f dispersion ($C_{100Hz}/C_{1MHz}$), and (b) leakage current ($|\log(I_{leakage})|$). Input parameters are ranked vertically based on their magnitude of effect on the model's outcome.

## Conclusions

In summary, we have constructed a framework for incorporating human input into multi-objective Bayesian optimization and have applied it to the optimization of $Al_2O_3$ sol-gel dielectric processed by photonic curing for neuromorphic electronics. This process shows a clear benefit over multi-objective optimization without human input. Furthermore, the resulting Pareto frontier in the final model comprises a range of processing conditions that will enable tunable design for fabricating neuromorphic transistors. Incorporating humans into the loop with machine learning models not only benefits experiments that suffer from a high failure rate, but it can also greatly accelerate optimization in systems where domain expertise from humans provides information that is difficult or time-consuming for a machine to obtain.

## Author contributions

**B. D.**: formal analysis, investigation, visualization, writing – original, and writing – review & editing. **J. M.-A.**: formal analysis, investigation, visualization, and writing – original. **A. T.**: methodology and writing – review & editing. **M. L.**: methodology, software, and writing – review & editing. **J. W. P. H.**: conceptualization, funding acquisition, supervision, and writing – review & editing.

## Conflicts of interest

There are no conflicts to declare.

## Data availability

Data supporting this article are available within the ESI† and GitHub repository: github.com/UTD-Hsu-Lab/ParetoUCB

## Acknowledgements

We thank Energy Materials Corporation for providing the PET substrates. This material is based upon work supported by the Air Force Office of Scientific Research under award number FA2386-24-1-4041 and by the National Science Foundation Award CMMI-2135203. J. M. gratefully acknowledges the support of the Program Fulbright García Robles fellowship. A.T. acknowledges funding from the European Union's Horizon 2020 research and innovation programme under the Marie Sklodowska-Curie grant agreement No 01059891. J.W.P.H. acknowledges the support of the Texas Instruments Distinguished Chair in Nanoelectronics.

## References

1. D. R. Muir and S. Sheik, *Nat Commun*, 2025, **16**, 3586.
2. A. Mehonic, *et al.*, *APL Mater*, 2024, **12**, 109201.
3. H. L. Park, Y. Lee, N. Kim, D. G. Seo, G. T. Go and T. W. Lee, *Advanced Materials*, 2020, **32**, 1903558.
4. H. Jang, J. Lee, C. J. Beak, S. Biswas, S. H. Lee and H. Kim, *Advanced Materials*, 2025, **37**, 2416073.
5. J. S. Lee, S. Lee and T. W. Noh, *Appl Phys Rev*, 2015, **2**, 031303.
6. Y. C. Mi, C. H. Yang, L. C. Shih and J. S. Chen, *Adv Opt Mater*, 2023, **11**, 2300089.
7. K. T. Chen, L. C. Shih, S. C. Mao and J. S. Chen, *ACS Appl Mater Interfaces*, 2022, **15**, 9593–9603.
8. S. Bolat, G. Torres Sevilla, A. Mancinelli, E. Gilshtein, J. Sastre, A. Cabas Vidani, D. Bachmann, I. Shorubalko, D. Briand, A. N. Tiwari and Y. E. Romanyuk, *Sci Rep*, 2020, **10**, 16664.
9. W. Xu, H. Wang, F. Xie, J. Chen, H. Cao and J. Bin Xu, *ACS Appl Mater Interfaces*, 2015, **7**, 5803–5810.
10. T. B. Daunis, J. M. H. Tran and J. W. P. Hsu, *ACS Appl Mater Interfaces*, 2018, **10**, 39435–39440.
11. T. B. Daunis, K. A. Schroder and J. W. P. Hsu, *npj Flexible Electronics*, 2020, **4**, 7.
12. K. A. Schroder, *Nanotechnology*, 2011, **2**, 220–223.
13. B. Shahriari, K. Swersky, Z. Wang, R. P. Adams and N. De Freitas, *Proceedings of the IEEE*, 2016, **104**, 148–175.
14. W. Xu, Z. Liu, R. T. Piper and J. W. P. Hsu, *Solar Energy Materials and Solar Cells*, 2023, **249**, 112055.
15. M. Kim, M. Y. Ha, W. Bin Jung, J. Yoon, E. Shin, I. doo Kim, W. B. Lee, Y. J. Kim and H. tae Jung, *Advanced Materials*, 2022, **34**, 2108900.
16. D. Xue, P. V. Balachandran, J. Hogden, J. Theiler, D. Xue and T. Lookman, *Nat Commun*, 2016, **7**, 1–9.
17. A. G. Kusne, H. Yu, C. Wu, H. Zhang, J. Hattrick-Simpers, B. DeCost, S. Sarker, C. Oses, C. Toher, S. Curtarolo, A. V. Davydov, R. Agarwal, L. A. Bendersky, M. Li, A. Mehta and I. Takeuchi, *Nat Commun*, 2020, **11**, 5966.
18. C. B. Wahl, M. Aykol, J. H. Swisher, J. H. Montoya, S. K. Suram and C. A. Mirkin, *Sci Adv*, 2021, **7**, 5505.
19. S. Daulton, M. Balandat and E. Bakshy, in *Advances in neural information processing systems*, 2021, vol. 34, pp. 2187–2200.
20. I. Couckuyt, D. Deschrijver and T. Dhaene, in *2012 IEEE Congress on Evolutionary Computation*, 2012, pp. 1–8.
21. O. Mamun, M. Bause and B. S. M. Ebna Hai, *Mach Learn Sci Technol*, 2025, **6**, 015001.
22. P. Ngatchou, A. Zarei and M. A. El-Sharkawi, in *Proceedings of the 13th International Conference on, Intelligent Systems Application to Power Systems*, 2005, pp. 84–91.
23. A. Tiihonen, L. Filstroff, P. Mikkola, E. Lehto, S. Kaski, M. T. Todorović and P. Rinke, in *AI for Accelerated Materials Design NeurIPS 2022 Workshop*, 2022.
24. Z. Liu, N. Rolston, A. C. Flick, T. W. Colburn, Z. Ren, R. H. Dauskardt and T. Buonassisi, *Joule*, 2022, **6**, 834–849.
25. novacentrix.com/documents/PF-Invent-Brochure.pdf
26. M. Konakovic Lukovic, Y. Tian and W. Matusik, in *Advances in Neural Information Processing Systems*, 2020, vol. 33, pp. 17708–17720.
27. M. Balandat, B. Karrer, D. Jiang, S. Daulton, B. Letham, A. G. Wilson and E. Bakshy, *Adv Neural Inf Process Syst*, 2020, **33**, 21524–21538.
28. S. Salgia, S. Vakili and Q. Zhao, in *Advances in Neural Information Processing Systems*, 2021, vol. 34, pp. 28836–28847.
29. A. K. Y. Low, F. Mekki-Berrada, A. Gupta, A. Ostudin, J. Xie, E. Vissol-Gaudin, Y. F. Lim, Q. Li, Y. S. Ong, S. A. Khan and K. Hippalgaonkar, *NPJ Comput Mater*, 2024, **10**, 104.
30. Y. K. Wakabayashi, T. Otsuka, Y. Krockenberger, H. Sawada, Y. Taniyasu and H. Yamamoto, *NPJ Comput Mater*, DOI:10.1038/s41524-022-00859-8.
31. K. Sattari, Y. Wu, Z. Chen, A. Mahjoubnia, C. Su and J. Lin, *Addit Manuf*, 2024, **86**, 104204.
32. R. J. Hickman, G. Tom, Y. Zou, M. Aldeghi and A. Aspuru-Guzik, *Digital Discovery*, 2025, **4**, 2104.
33. S. Lundberg and S.-I. Lee, in *Advances in Neural Information Processing Systems*, 2017, vol. 30.

# Multi-objective Bayesian Optimization with Human-in-the-Loop for Flexible Neuromorphic Electronics Fabrication

# Supplementary Information

Benius Dunn,[a] Javier Meza-Arroyo,[b] Armi Tiihonen,[b] Mark Lee [c] and Julia W. P. Hsu *[ac]

[a] Department of Materials Science and Engineering, The University of Texas at Dallas, Richardson, Texas 75080, USA

[b] Department of Applied Physics, Aalto University, Espoo, Finland

[c] Department of Physics, The University of Texas at Dallas, Richardson, Texas 75080, USA

*Corresponding author.  Email: jwhsu@utdallas.edu

# Initial Latin hypercube sampling (LHS) photonic curing conditions

Table S1. Ranges and step sizes for LHS sampling

| Variable Name | Minimum Value | Maximum Value | Step Size | Number of steps |
|---|---|---|---|---|
| Radiant Energy (J/cm$^2$) | 1.0 | 7.0 | 0.2 | 31 |
| Pulse Count | 1 | 20 | 1 | 20 |
| Pulse Length (ms) | 1 | 20 | 1 | 20 |
| Number of Micropulses | 1 | 30 | 1 | 30 |
| Duty Cycle (%) | 20 | 70 | 5 | 11 |

Table S2. Photonic curing input parameters, human observation scores, and measured device outputs for initial LHS conditions. Data for all conditions is available through UTD-Hsu-Lab/ParetoUCB.

| Condition # | Radiant Energy (J/cm$^2$) | Pulse Count | Pulse Length (ms) | Micropulse Count | Duty Cycle (%) | Pulse Voltage (V) | Conversion Score (-1 – 1) | $C_{100Hz}/C_{1MHz}$ | $|\log(I_{leakage})|$ |
|---|---|---|---|---|---|---|---|---|---|
| 1 | 1.4 | 5 | 19 | 22 | 45 | 216 | -1.0 | - | - |
| 2 | 3.8 | 8 | 5 | 12 | 35 | 447 | 1.0 | - | - |
| 3 | 3.2 | 4 | 4 | 19 | 30 | 492 | 1.0 | - | - |
| 4 | 6 | 14 | 3 | 13 | 20 | 810 | 1.0 | - | - |
| 5 | 5.4 | 9 | 19 | 17 | 40 | 374 | 1.0 | - | - |
| 6 | 4.4 | 17 | 10 | 6 | 25 | 418 | 1.0 | - | - |
| 7 | 3 | 7 | 17 | 7 | 25 | 323 | 1.0 | - | - |
| 8 | 6.6 | 12 | 2 | 15 | 50 | 651 | 1.0 | - | - |
| 9 | 2.4 | 1 | 13 | 1 | 65 | 256 | -1.0 | - | - |
| 10 | 2 | 18 | 8 | 10 | 45 | 290 | -0.5 | 2.00 ± 0.23 | 1.09 ± 0.69 |
| 11 | 6.8 | 17 | 2 | 5 | 30 | 766 | 1.0 | - | - |
| 12 | 3.4 | 6 | 12 | 9 | 40 | 330 | 1.0 | - | - |
| 13 | 1 | 14 | 14 | 25 | 60 | 190 | -1.0 | - | - |
| 14 | 6.2 | 20 | 6 | 16 | 55 | 450 | 1.0 | - | - |
| 15 | 1.6 | 3 | 15 | 4 | 55 | 220 | -1.0 | - | - |
| 16 | 4.8 | 16 | 7 | 26 | 65 | 385 | 0.5 | 1.19 ± 0.08 | 4.90 ± 0.27 |
| 17 | 4.2 | 4 | 9 | 23 | 70 | 344 | 0.0 | 1.58 ± 0.20 | 3.96 ± 0.38 |
| 18 | 2.8 | 10 | 11 | 29 | 50 | 305 | -0.5 | 2.10 ± 0.40 | 3.55 ± 0.22 |
| 19 | 5.8 | 11 | 18 | 27 | 60 | 371 | 0.5 | 2.30 ± 0.43 | 3.46 ± 0.43 |
| 20 | 5 | 13 | 15 | 21 | 35 | 388 | 0.0 | 2.27 ± 0.43 | 3.07 ± 0.23 |
| 21 | 3.8 | 6 | 10 | 27 | 25 | 426 | 0.0 | 2.36 ± 0.49 | 3.44 ± 0.40 |
| 22 | 3.2 | 7 | 12 | 21 | 45 | 322 | -0.5 | 3.30 ± 0.42 | 3.30 ± 0.17 |
| 23 | 5.8 | 9 | 17 | 10 | 40 | 388 | 1.0 | - | - |
| 24 | 1.8 | 2 | 8 | 23 | 50 | 271 | -1.0 | - | - |
| 25 | 5.2 | 6 | 10 | 30 | 55 | 388 | 0.5 | 2.00 ± 0.29 | 0.88 ± 0.97 |
| 26 | 2.6 | 10 | 8 | 15 | 70 | 288 | 0.0 | 2.50 ± 0.40 | 3.88 ± 0.23 |
| 27 | 1.8 | 8 | 4 | 25 | 30 | 412 | -1.0 | - | - |
| 28 | 2.4 | 4 | 20 | 17 | 65 | 252 | -1.0 | - | - |
| 29 | 1.2 | 3 | 3 | 5 | 35 | 346 | -1.0 | - | - |
| 30 | 4.4 | 3 | 18 | 3 | 20 | 382 | 1.0 | - | - |

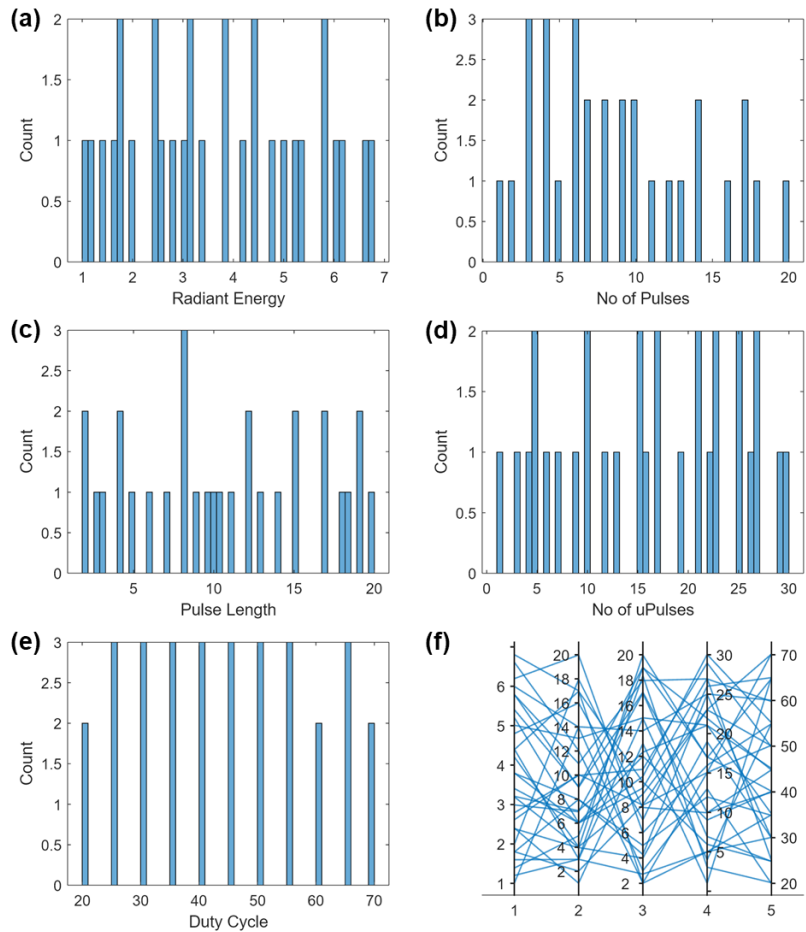

**Fig. S1.** Histograms showing distribution of initial LHS photonic curing parameters (a) radiant energy, (b) number of pulses, (c) pulse length, (d) number of micropulses, and (e) duty cycle. (f) Parallel plot for all 30 initial LHS conditions across all five input parameters to demonstrate exploration across the entire input space.

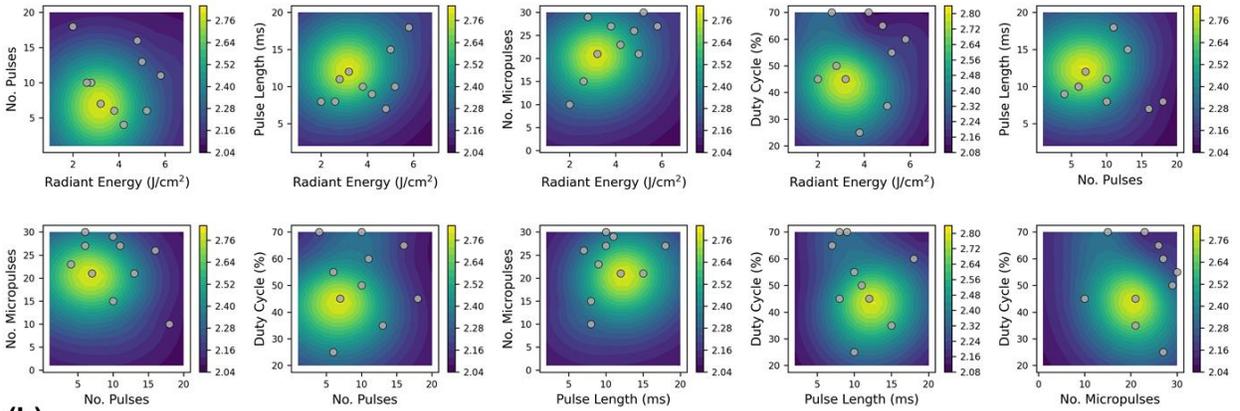
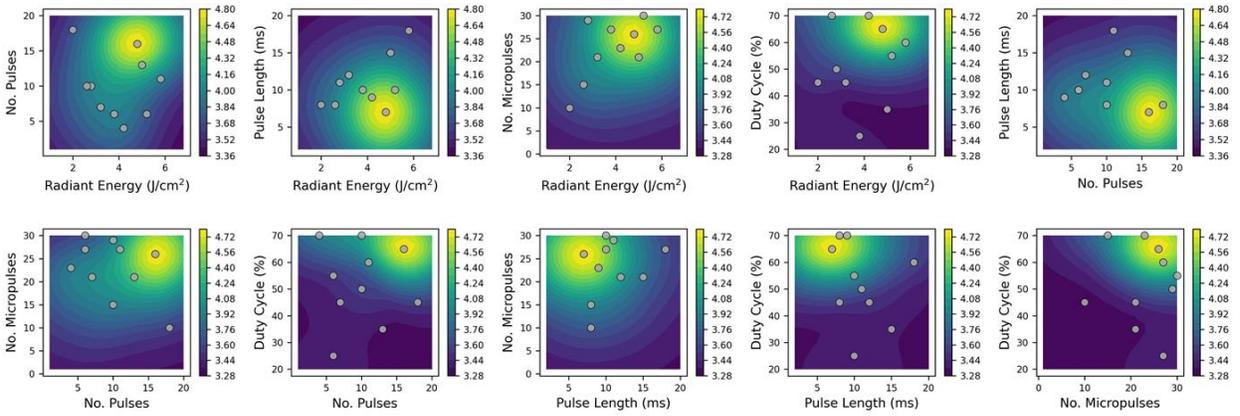

**Fig. S2.** GPR models built from the data collected on functional devices made with initial 30 LHS conditions: (a) C-f dispersion ($C_{100Hz}/C_{1MHz}$), and (b) leakage current ($|\log(I_{leakage})|$).

# Human-in-the-Loop methodology

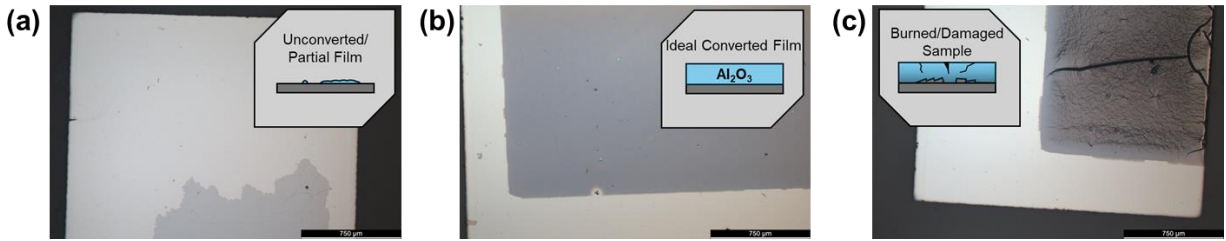

**Fig. S3.** Optical microscope images of films after photonic curing showing different conversion outcomes: (a) under-converted film with partial film remaining, (b) good conversion that can result in functional devices, and (c) burned sample.

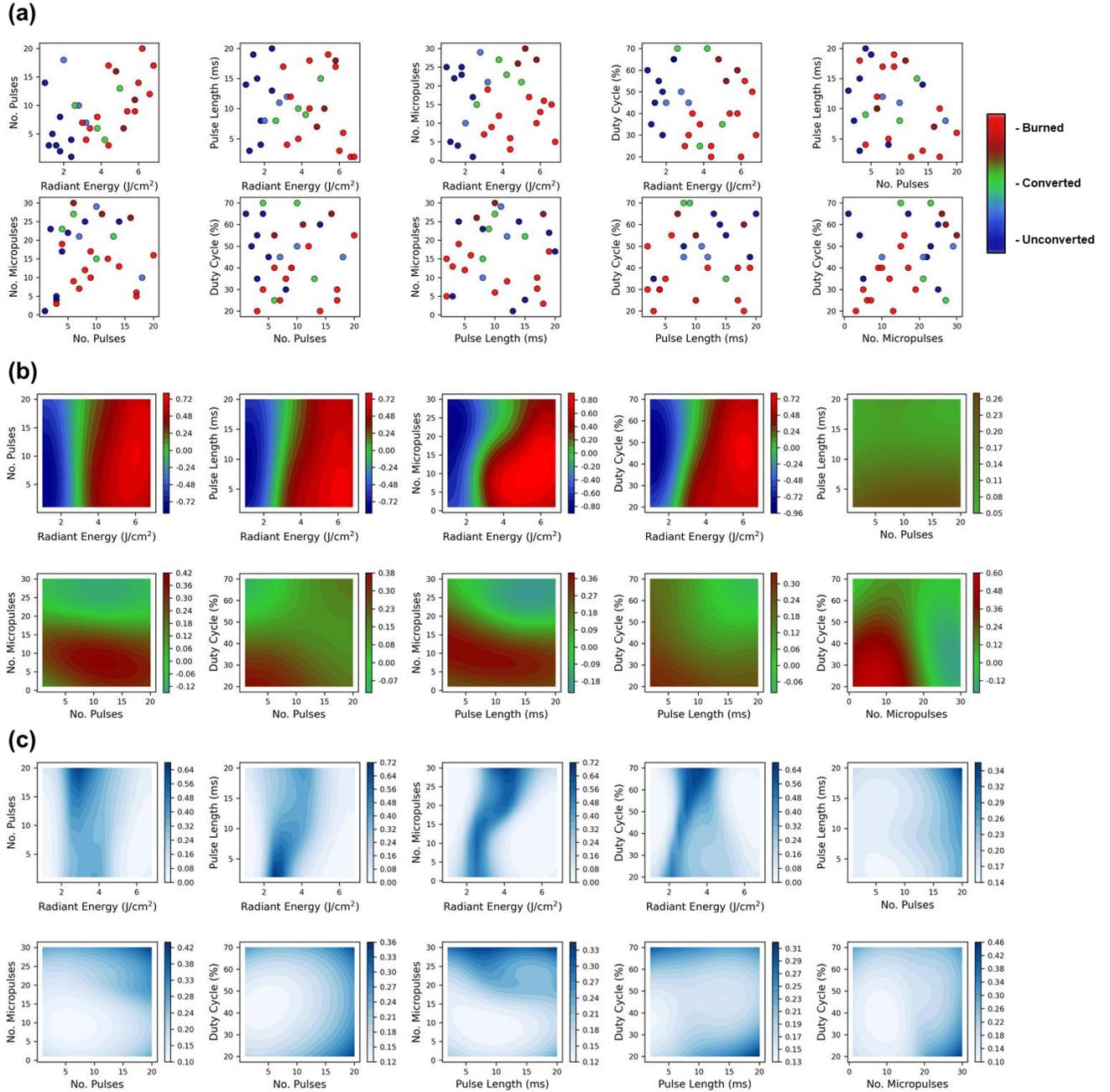

**Fig. S4.** (a) Human observation scores of the initial 30 LHS conditions, (b) GPR model trained on human observation scores, and (c) resulting HITL probability distribution $P_{constraint}$ obtained from a Gaussian transformation of (b).

**Dominated hypervolume evolution & final GPR models' accuracy**

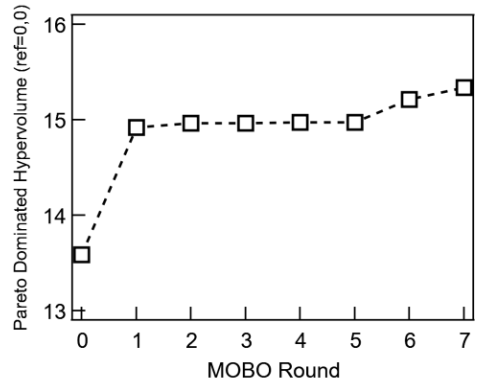

**Fig. S5.** Dominated hypervolume by Pareto optimal data points across each round of MOBO, starting with initial LHS (round 0)

When using a machine learning model as a surrogate for real experimental data, fitting is often less ideal than with synthetic data due to sample-to-sample variation and limited availability of data points for model training. As depicted in Fig. S6(a), after completing all MOBO rounds, the final GPR model for C-f dispersion tends to overestimate values of $C_{100Hz}/C_{1MHz}$ for values greater than ~2. However, the region of feasible device fabrication is limited to values of $C_{100Hz}/C_{1MHz}$ less than 2, for which the model shows suitable accuracy. The final GPR model of leakage current similarly shows reasonable accuracy for the region of feasible device fabrication with values of $|\log(I_{leakage})| \geq 4$ in Fig. S6(b). The C-f dispersion model has a linear fit slope of 0.723 with a $R^2$ value of 0.938. The leakage current model has a linear fit slope of 0.852 with a $R^2$ value of 0.964.

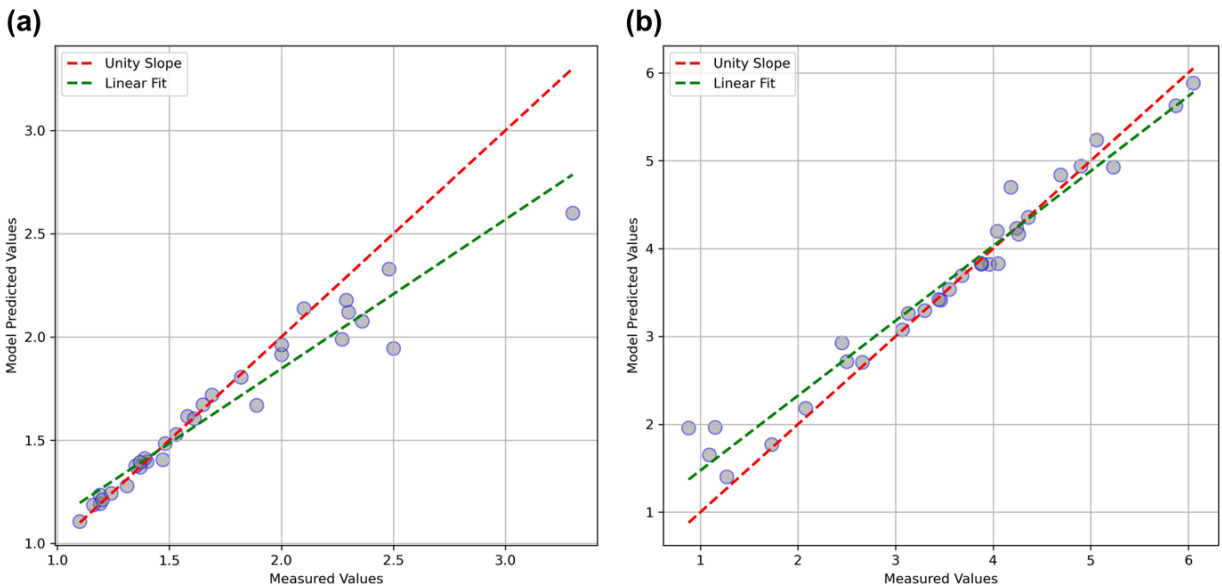

**Fig. S6.** Final GPR model predictions versus measured output values for (a) C-f dispersion ($C_{100Hz}/C_{1MHz}$), and (b) leakage current ($|\log(I_{leakage})|$).

# Physical and chemical property comparison of dielectrics from MOBO conditions #40 and #66

Conditions #40 and #66 represent the two extreme conditions along the Pareto frontier within the realm of feasible device fabrication ($I_{leakage} < 10^{-4}$ A/cm$^2$). This section compares the processing conditions and the physical and chemical properties of Al$_2$O$_3$ dielectrics made using these two conditions. Fig. S7 shows simulated temperature profiles over time for a single photonic curing pulse under conditions #40 and #66. The temperature profiles are constructed using the SimPulse® photonic curing simulation software, which is included with the PulseForge Invent system. Samples are modeled as a material stack that consists of 100 µm PET, 100 nm aluminum metal, and 70 nm Al$_2$O$_3$. Profiles are generated in surface absorption mode with 11% surface absorption, which is determined using a NIST-calibrated bolometer to estimate transmission % and reflectance mode UV-vis to estimate reflectance %. Profiles for the top of the Al$_2$O$_3$ in Fig. S7(a) show the sample processed with condition #40 reaching a peak temperature of 159°C, while the sample under condition #66 reaches a lower peak temperature of 127°C. For both conditions, the bottom of the PET substrate does not exceed 55°C, as shown in Fig. S7(b). While the one-dimensional thermal model is very simple, the difference in peak temperature of the film suggests that condition #40 induces a greater conversion from precursor to oxide dielectric.[1]

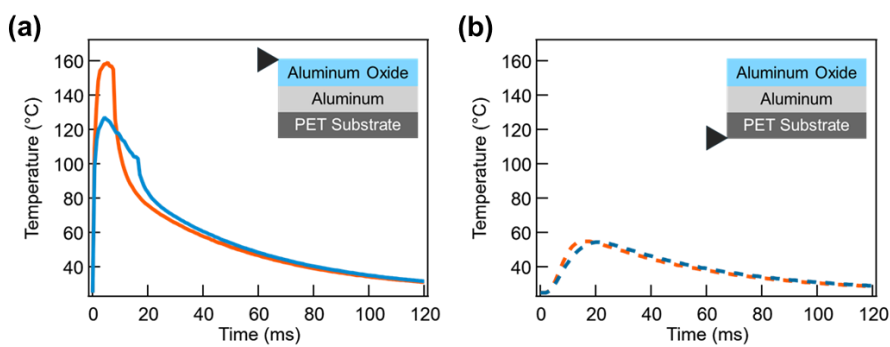

**Fig. S7.** SimPulse® simulated temperature profiles for a single pulse of MOBO conditions #40 (orange) and #66 (blue) (a) on the Al$_2$O$_3$ film surface and (b) at the bottom of the PET substrate.

We further investigate the chemical difference between the films after photonic curing by X-ray photoelectron spectroscopy (XPS) to study the conversion outcome. XPS is performed on a Ulvac-PHI VersaProbe2 with a monochromated Al Kα source (1486.8 eV) at an angle of 45° to the sample surface. O1s spectra are averaged over 20 scans with an energy step of 0.2 eV and a pass energy of 23.5 eV. N1s spectra are averaged over 100 scans with an energy step of 0.2 eV and a pass energy of 187.85 eV. Fitting is performed with the CasaXPS software using a mixed 30:70 Gaussian:Lorenzian functions for all peaks, and binding energy is calibrated using the 284.8 eV C1s peak from adventitious carbon.

O1s spectra are fit to two peaks reflecting the different chemical environments of oxygen within the dielectric: 530.4 eV peak corresponding to metal oxide and 532 eV peak corresponding to metal hydroxide. Dielectric fabricated using condition #40 shows a metal oxide signal of 19% (Fig. S8(a)) while the one fabricated using condition #66 only has a metal oxide signal of 6.7% (Fig. S8(b)), in line with the expectation that the film reaches a higher temperature under condition #40 and has a greater conversion to metal oxide.

Also shown in Fig. S8(c) is the N1s signal intensity at 407 eV for each sample. This N1s signal arises primarily from residual nitrate ($NO_3^-$) from the aluminum nitrate precursor. As such, the lower N1s signal present from condition #40 provides further evidence of greater conversion from precursor to metal oxide. This analysis is consistent with the expected behavior of the simulated temperature profiles in Fig. S7.

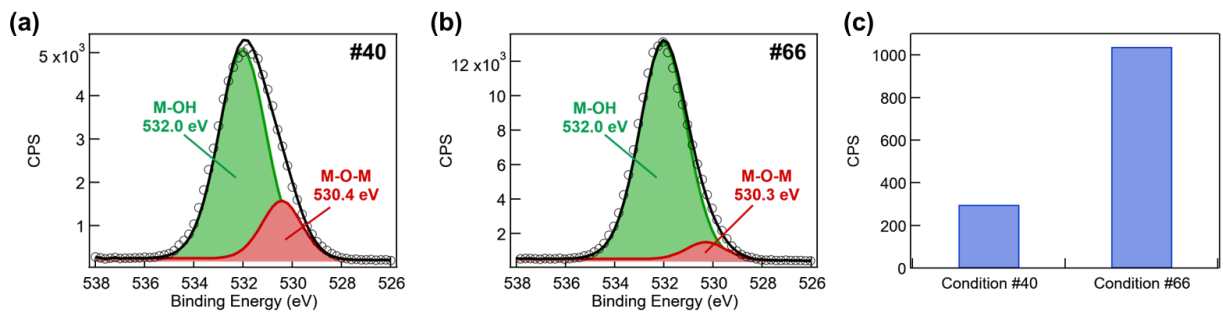

**Fig. S8.** XPS O1s spectra for (a) MOBO condition #40 and (b) #66. (c) XPS N1s peak intensity comparison for conditions #40 and #66.

# HITL methodology comparison data and summary of methods used for each round of MOBO

The acquisition functions used in each round of MOBO, as well as whether the HITL methodology is implemented, are indicated in Table S4. Rounds 1a and 1b are both picked by the models trained on the 10 successful devices from the 30 LHS conditions only, their only difference being the acquisition function used. Thus, round 2 begins once the model is updated with the data from rounds 1a and 1b. Rounds 1′ and 2′ are the picks by models constrained with HITL knowledge (Fig. 4(c)). Round 1′ is trained only on the results of the initial 30 LHS conditions, while round 2′ is trained on the updated model with data from LHS and round 1′.

**Table S4.** Conditions for each round of multi-objective optimization: acquisition function used (qEHVI: parallel expected hypervolume improvement; ParUCB: Pareto-UCB batch picking), and whether human-in-the-loop (HITL) is used in each round.

| Round # | 1a | 1b | 2 | 3 | 4 | 5 | 6 | 7 |
|---|---|---|---|---|---|---|---|---|
| Acquisition Function | qEHVI | ParUCB | qEHVI | qEHVI | qEHVI | qEHVI | ParUCB | ParUCB |
| HITL | ✗ | ✗ | ✗ | ✓ | ✓ | ✓ | ✓ | ✓ |

| Round # | 1′ | 2′ |
|---|---|---|
| Acquisition Function | ParUCB | ParUCB |
| HITL | ✓ | ✓ |

**Table S5.** Photonic curing input parameters, human observation scores, and measured device outputs for rounds of HITL methodology comparison.

| Round # | Condition # | Radiant Energy (J/cm$^2$) | Pulse Count | Pulse Length (ms) | Micropulse Count | Duty Cycle (%) | Pulse Voltage (V) | Conversion Score (-1 – 1) | $C_{100Hz}/C_{1MHz}$ | $|\log(I_{leakage})|$ |
|---|---|---|---|---|---|---|---|---|---|---|
| 1a | 31 | 2.7 | 8 | 11 | 18 | 54 | 294 | -1.0 | - | - |
| 1a | 32 | 3.1 | 6 | 13 | 21 | 37 | 328 | -1.0 | - | - |
| 1a | 33 | 3.4 | 12 | 8 | 20 | 70 | 322 | 0.0 | 1.53 ± 0.15 | 2.50 ± 0.39 |
| 1a | 34 | 2.4 | 5 | 8 | 14 | 70 | 280 | -1.0 | - | - |
| 1a | 35 | 3.3 | 9 | 14 | 21 | 47 | 315 | -1.0 | - | - |
| 1b | 36 | 2.8 | 8 | 11 | 20 | 52 | 301 | -1.0 | - | - |
| 1b | 37 | 3 | 8 | 11 | 21 | 45 | 320 | -1.0 | - | - |
| 1b | 38 | 4 | 14 | 8 | 21 | 70 | 345 | 0.0 | 1.19 ± 0.12 | 5.23 ± 0.50 |
| 1b | 39 | 4.3 | 14 | 8 | 22 | 67 | 357 | 0.0 | 1.16 ± 0.15 | 5.87 ± 0.30 |
| 1b | 40 | 4.5 | 16 | 7 | 23 | 70 | 369 | 0.5 | 1.10 ± 0.04 | 6.05 ± 0.24 |
| 2 | 41 | 2.4 | 6 | 12 | 20 | 43 | 292 | -1.0 | - | - |
| 2 | 42 | 4 | 7 | 12 | 20 | 43 | 354 | -1.0 | - | - |
| 2 | 43 | 4.4 | 17 | 11 | 21 | 65 | 346 | 0.0 | 1.40 ± 0.07 | 4.18 ± 1.05 |
| 2 | 44 | 3 | 4 | 13 | 26 | 39 | 322 | -1.0 | - | - |
| 2 | 45 | 2.1 | 6 | 9 | 12 | 66 | 262 | -1.0 | - | - |
| 1′ | 31′ | 3.7 | 14 | 13 | 20 | 70 | 314 | 0.0 | 2.76 ± 0.61 | 3.44 ± 0.13 |
| 1′ | 32′ | 2.9 | 1 | 13 | 18 | 37 | 320 | -1.0 | - | - |
| 1′ | 33′ | 3.3 | 15 | 9 | 26 | 70 | 315 | 0.0 | 2.93 ± 0.36 | 2.99 ± 0.27 |
| 1′ | 34′ | 3.3 | 6 | 12 | 18 | 40 | 336 | 0.0 | 1.98 ± 0.22 | 3.44 ± 0.80 |
| 1′ | 35′ | 3.7 | 10 | 15 | 15 | 67 | 311 | 0.0 | 2.60 ± 0.46 | 3.24 ± 0.14 |
| 2′ | 36′ | 2.9 | 6 | 6 | 20 | 70 | 321 | 0.0 | 2.09 ± 0.16 | 3.23 ± 0.73 |
| 2′ | 37′ | 3.3 | 11 | 10 | 18 | 60 | 319 | 0.0 | 3.15 ± 0.69 | 3.13 ± 0.25 |
| 2′ | 38′ | 4.5 | 17 | 13 | 28 | 57 | 352 | 0.0 | 1.46 ± 0.11 | 4.30 ± 0.24 |
| 2′ | 39′ | 2.9 | 9 | 7 | 20 | 63 | 320 | 0.0 | 2.52 ± 0.34 | 3.29 ± 0.13 |
| 2′ | 40′ | 3.3 | 15 | 10 | 20 | 63 | 316 | 0.0 | 2.23 ± 0.27 | 3.04 ± 0.60 |

## SHAP analysis scatter plots for individual input variables

The correlation of SHAP values for photonic curing parameters versus input values are depicted by scatter plots for the C-f dispersion model (Fig. S9) and leakage current model (Fig. S10). A positive correlation between SHAP value and input parameter value, such as for pulse length in Fig. S9(c), indicates that increasing this input will increase the output of the model. Conversely, a negative correlation between SHAP value, such as for radiant energy in Fig. S9(a), indicates that increasing this input will decrease the output of the model.

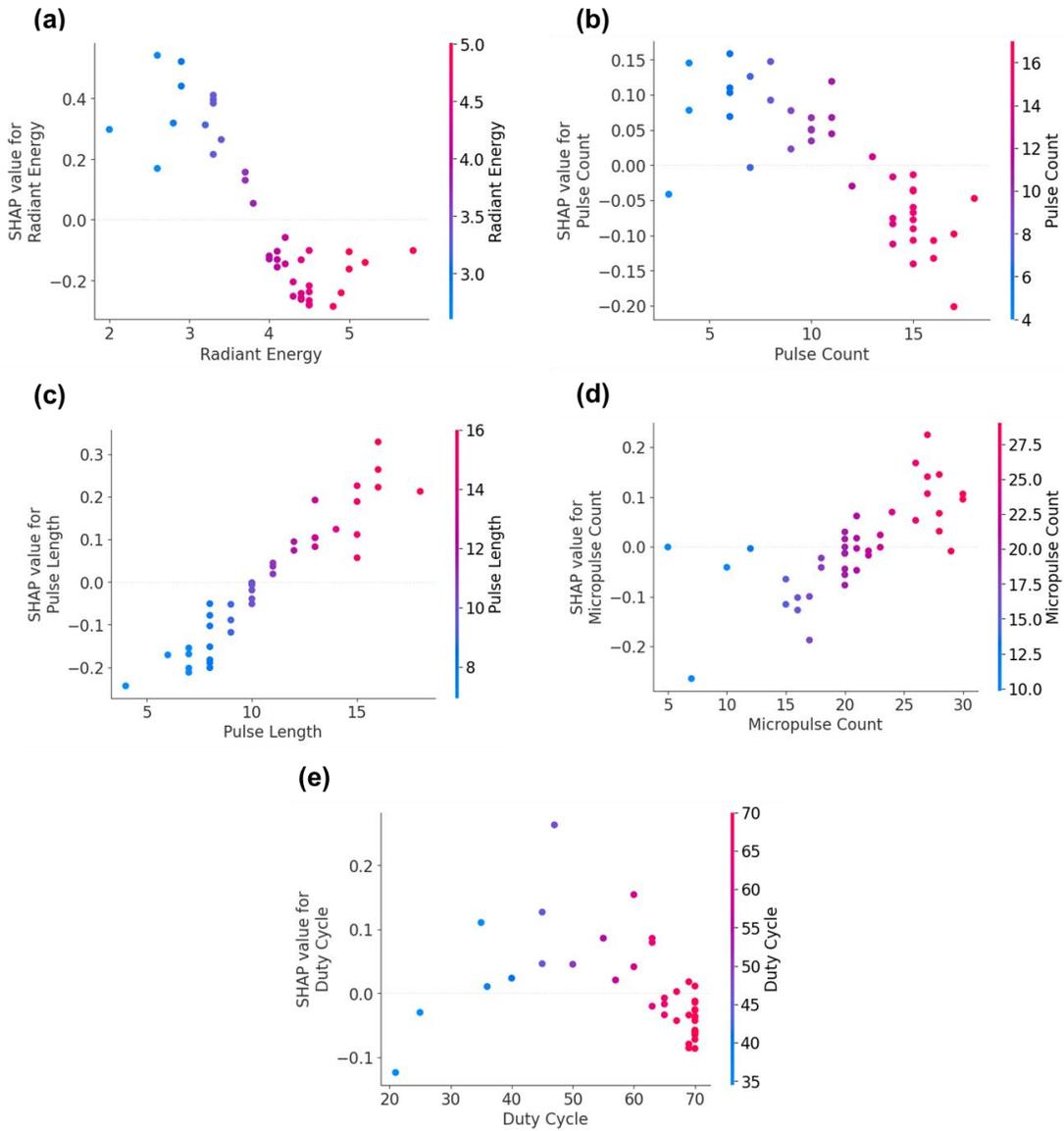

**Fig. S9.** SHAP values of C-f dispersion ($C_{100Hz}/C_{1MHz}$) model versus input values for (a) radiant energy, (b) pulse count, (c) pulse length, (d) micropulse count, and (e) duty cycle.

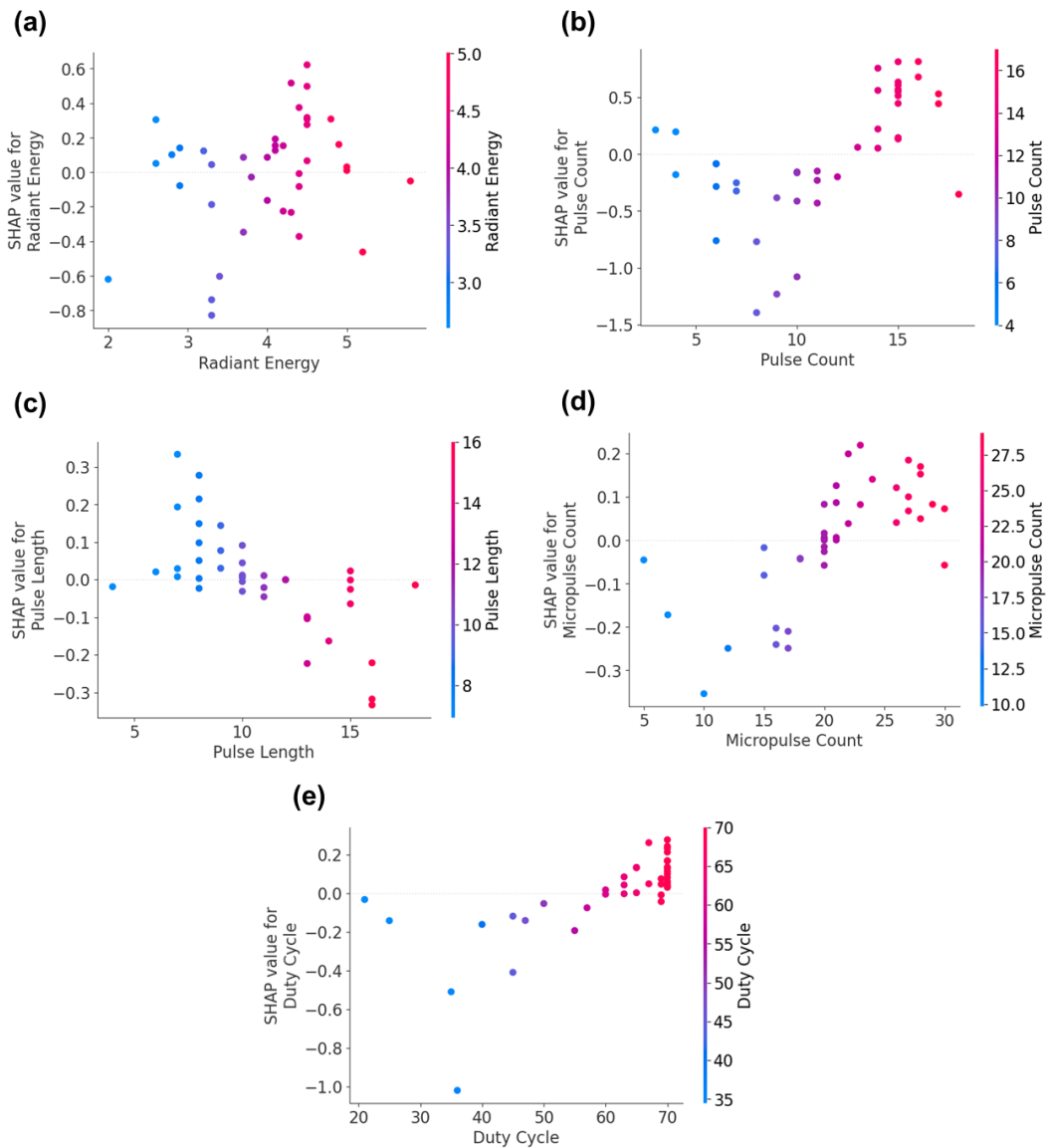

**Fig. S10.** SHAP values of leakage current ($|\log(I_{leakage})|$) model versus feature values for (a) radiant energy, (b) pulse count, (c) pulse length, (d) micropulse count, and (e) duty cycle.

## References

1   E. A. Cochran, K. N. Woods, D. W. Johnson, C. J. Page and S. W. Boettcher, *J Mater Chem A Mater*, 2019, **7**, 24124–24149.